\documentclass[apjl]{emulateapj}

\slugcomment{ApJ Letters - Received 2005 June 20; accepted 2005 August 25}
\shorttitle{The light of MgII absorbing galaxies}
\shortauthors{S. Zibetti et al.}

\newcommand{\MgII}{MgII}
\newcommand{\be}{\begin{equation}}
\newcommand{\ee}{\end{equation}}
\newcommand{\bea}{\begin{eqnarray}}
\newcommand{\eea}{\end{eqnarray}}

\begin{document}

\title{Constraining the photometric properties of MgII absorbing galaxies with the SDSS}

\author{Stefano Zibetti \altaffilmark{1},
  Brice M\'enard         \altaffilmark{2},
  Daniel Nestor           \altaffilmark{3}, and
  David Turnshek          \altaffilmark{4}}
\altaffiltext{1}{Max-Planck-Institut f\"ur Extraterrestrische Physik, Gie\ss enbachstra\ss e,
D-85748,Garching bei M\"unchen, Germany, e-mail szibetti@mpe.mpg.de}
\altaffiltext{2}{Institute for Advanced Study, Einstein Drive, Princeton NJ
  08540, USA, e-mail menard@ias.edu} 
\altaffiltext{3}{Dept. of Astronomy, University of Florida, Gainesville, FL 32611, USA,
e-mail dbn@astro.ufl.edu}   
\altaffiltext{4}{Dept. of Physics and Astronomy, University of Pittsburgh,
  Pittsburgh, PA 15260, USA, e-mail turnshek@pitt.edu} 

\begin{abstract}
Using a sample of nearly $700$ quasars with strong ($\mathrm{W_0(2796)}>0.8\;\mathrm{\AA}$)
MgII absorption
lines detected in the Early Data Release of the
SDSS, we demonstrate the feasibility of measuring the photometric
properties of the absorber systems by stacking SDSS imaging data.
As MgII lines can be observed in the range
$0.37<z_{\mathrm{abs}}<2.2$, the absorbing galaxies are in general
not identified in SDSS images, but they produce systematic light
excesses around QSOs which can be detected with a statistical
analysis.
In this \emph{Letter} we present a $6~\sigma$ detection
of this effect over the whole sample in $i$-band, rising to
$9.4~\sigma$ for a low-redshift subsample with $0.37<z_{\mathrm{abs}}\leq0.82$.
We use a
control sample of QSOs without strong MgII absorption lines to quantify and
remove systematics with typical 10-20\% accuracy. The signal varies as
expected as a function of absorber redshift. For the low $z_{\mathrm{abs}}$
subsample we can reliably estimate the average
luminosities per MgII absorber system in the $g$, $r$, and $i$ bands
and find them to be compatible with a few-hundred-Myr old stellar 
population of $M_r\sim -21$ in the rest frame.
Colors are also consistent with typical absorbing galaxies
resembling local Sb-c spirals.
Our technique does not require any spectroscopic follow-up and does 
not suffer from confusion with other galaxies arising along the
line-of-sight. It will be applied to larger samples and other line
species in upcoming studies.
\end{abstract}


\section*{Introduction}
\label{intro_section}

A number of studies have inferred and then shown that strong metal
absorption lines in a quasar spectrum are due to gaseous
clouds associated to galaxies close to the line-of-sight
(e.g. \citealt{Bahcall69,Bergeron91,Steidel94}). However, the systematic
characterization of the emission properties of these absorber
systems is still limited, especially at high redshifts (see for
example \citealt{Churchill05}). In order to use
absorption-line measurements for investigating galaxy formation and
evolution, characterizing the underlying population of galaxies seen
in absorption is an important task to achieve.

Among metal absorption lines, the \MgII\ doublet
$\lambda\lambda2796,2803$, detectable from the ground at $0.2<z<2.2$,
has been widely used. It arises in gas spanning more than five decades
of neutral hydrogen column density (e.g. \citealt{Churchill00}) and
the strong systems, defined by $W_0(\lambda\lambda2796)\gtrsim0.3$
\AA\ are found in the vicinity of galaxies spanning a large range of
morphologies. Previous attempts to constrain the properties of MgII
absorbers have relied on deep imaging and spectroscopic follow-up of
(arbitrarily faint) galaxies in the quasar field in order to identify
a potential galaxy responsible for the absorption seen in the quasar
spectrum.  Such expensive studies have been limited to samples of a
few tens of quasars.

Here we propose a new approach to measure the systematic photometric
properties of large samples ($N\gg 100$) of absorbing systems, which
combines statistical analysis of both spectroscopic and imaging
datasets of the Sloan Digital Survey (hereafter SDSS,
\citealt{York00}).  As MgII lines can only be detected in SDSS spectra in the
range $0.37\lesssim z\lesssim 2.2$,
the absorbing galaxies are not identified and, in general, only marginally
detected in individual SDSS images. However, they do
produce systematic light excesses around the background QSOs that can
be detected with a statistical analysis consisting of stacking a large
number of absorbed QSO images. Such a statistical approach has been
already proved to be very powerful in detecting and characterizing the
very low-surface brightness \emph{diffuse} light in galaxy halos 
\citep{Zibetti04} and the intracluster light \citep{Zibetti05}. In this
\emph{Letter} we demonstrate that significant detections can be obtained
with this technique also for localized sources (the absorbing galaxies)
in the presence of strong background contamination.
Preliminary results on the photometric properties of the low $z$
($<0.82$) absorbers are also analyzed. More detailed results and analysis
for a larger sample will be presented in a forthcoming paper.

The paper is organized as follows. In Section \ref{section_data} we
present the samples of QSOs used in our analysis. The image processing
and stacking are described in Sec. \ref{section_processing}. The
results are presented and analyzed in Sec. \ref{section_results} and
conclusive remarks and forthcoming developments are outlined in
Sec. \ref{section_conclusions}. Standard $\Lambda$CDM cosmology
($\Omega_m=0.3$, $\Omega_\Lambda=0.7$) and $H_0=70 \mathrm{~km~sec}^{-1}
\mathrm{~Mpc}^{-1}$ are adopted throughout the paper.

\section{The data}\label{section_data}
\subsection{The MgII absorber sample}

\begin{figure}
\plotone{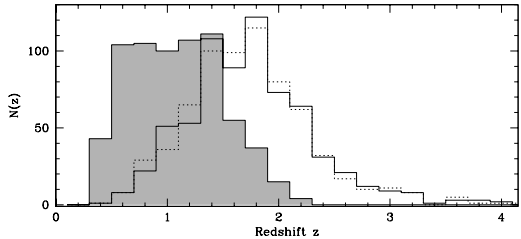}
\caption{Redshift distribution of MgII absorbers with
$W_0>0.8$\AA ~(shaded histogram), of their background quasars (solid line),
and of the \emph{reference} quasars (dotted line).
\label{plot_redshift}}
\end{figure}

We use the sample of \MgII\ absorption line systems compiled by 
\cite{Nestor04} and based on SDSS EDR data \citep{edr}. We focus
our analysis only on strong systems with a rest equivalent width
$\mathrm{W_0(2796)}>0.8\;\mathrm{\AA}$. In this range the detection
completeness is above 95\%. In this section we briefly summarize the main
steps involved in the absorption line detection procedure.  For
details on the quasar and absorber catalogs, we refer the reader to
\cite{Schneider02} and \cite{Nestor04}.\\  The SDSS EDR provides
approximately 3700 QSO spectra of sufficiently high redshift to  allow
the detection of intervening \MgII\ absorption lines.  The
continuum-normalized spectra were searched for \MgII\
$\lambda\lambda2796,2803$ doublets using automated routines and
interactive confirmation/rejection of candidate systems.  The rest
equivalent widths $W_0$ of confirmed systems were measured using an
optimal extraction method employing Gaussian line-profiles (or more
complex ones where appropriate).  All systems were
checked for blends and other special cases.
A 5$\sigma$ significance level was
required for all $\lambda2796$ lines, as well as a $3\sigma$
significance level for the corresponding $\lambda2803$ lines.  Only
systems 0.1c blue-ward of the quasar redshift and red-ward of
Ly$-\alpha$ emission 
(so as to avoid the Lyman-$\alpha$ forest) 
were accepted. Systems with separations of less
than 500 km/s were considered as a single system. 
In order to avoid confusion originating from the superposition of multiple
absorbers along the same line-of-sight, we exclude from our sample all QSOs
where distinct $\mathrm{W_0(2796)}>0.8\;\mathrm{\AA}$
absorber systems are detected: they represent 16.3\% of QSOs with 
$\mathrm{W_0(2796)}>0.8\;\mathrm{\AA}$ absorptions.
The final sample comprises 683 QSOs, out of which 2 are excluded because
they lie too close to bright stars. The absorber and QSO redshift distribution
are shown in Figure \ref{plot_redshift}.

\subsection{The reference quasar sample}
\label{section_selection}
In order to measure possible systematic light excess around QSOs
that could contaminate the signal originating from absorbing galaxies,
we use a control sample made of QSOs without strong absorbers.  
For each QSO with an absorber system, we randomly look for another QSO
without any detected absorber down to $W_0^{\lambda2796}=0.8\,\mathrm{\AA}$,
such that their redshifts are matched within 0.1\footnote{Only in one case
the best possible match is within 0.3.} and $i$-band brightnesses
differ by less than 0.5 mag. In
addition, we require that their spectra have similar signal-to-noise (S/N)
at the wavelengths
corresponding to the MgII absorption features. The last two conditions
also imply that the average overall S/N per pixel in the spectra 
of the two samples differs by less than 0.1.
Thus our procedure ensures that (i) the probability of contamination by
undetected MgII systems along the line-of-sight is the same for both 
QSOs and (ii) photometric systematics which may possibly arise from 
residuals of the PSF
subtraction (see Sec. \ref{PSFsub} and \ref{section_results}) affect both 
samples in the same way.
Point (i) has also been verified by means of Monte Carlo simulations. 
The QSOs selected in this way are called \emph{reference} QSOs
in the following. Their redshift distribution is also shown in Fig.
\ref{plot_redshift} (dotted histogram).

\section{Image stacking}\label{section_processing}

The image stacking technique aims at producing a high 
S/N \emph{average} image of the QSOs, possibly surrounded by
absorbing galaxies. It consists of four basic steps: (i) shifting to
make images superposable; (ii) intensity rescaling to a uniform
photometric calibration; (iii) masking of all unwanted sources; (iv)
average-combination of images, after rejecting masked regions.
All the procedures will be thoroughly described in a forthcoming paper.
Here we summarize the main points.

We use the raw ``corrected frames'' (fpC, bias and flat-field
corrected) publicly available in the Third Data Release of the SDSS
(\citealt{dr3}, DR3). Steps (i) and (ii) rely on the centroid calculations and
on the photometric calibrations provided by the DR3 catalogs and are
performed using standard IRAF tasks. A bi-cubic spline interpolant
is used to shift images.  Regarding step (iii), we note that an ideal
mask should retain only the QSO and the absorbing galaxy in each
image, while blanking out completely all other sources in order to
minimize the background noise. In practice we design our masking
algorithm to leave the QSO unmasked and to blank all sources which are
likely to be in the foreground with respect to the absorber.  For each QSO
we create three independent masks in $g$-, $r$- and $i$-band, that are
eventually combined in the final mask.
In each band, a first mask is obtained from the segmentation image
output by SExtractor \citep{bertin96} for a \emph{clean} image of the
field, from which we have subtracted the 
de-blended ``Atlas Image'' of the QSO produced by the \texttt{PHOTO}
pipeline \citep{Lupton01}.
This step ensures optimal de-blending of sources from the QSO and leaves
the QSO unmasked.
We retain the masks for all objects with a flux exceeding
that expected from \citeauthor{BC03}'s population synthesis
models (\citeyear{BC03}, hereafter BC03) for an extreme case of a 
luminous blue galaxy: a stellar
population with $\mathrm{M}_i=-22.0$ produced by a burst $10^7$ Gyr
long, observed immediately after the burst end at the redshift of the
absorber. Fainter objects are left unmasked. Our tests have shown that
the final results do not critically depend on the assumed
spectral energy distribution.  Finally, the masks in the three bands
are combined and conservative masks are applied to all objects 
classified as stars by \texttt{PHOTO} with $r<20.0$.

After subtracting the background value evaluated on the masked image
in an annulus between 200 and 250 pixels, the original images are shifted,
intensity-calibrated, and averaged, excluding all masked
pixels. Azimuthally averaged surface brightness (SB) profiles are extracted
in a series of circular apertures which are optimized to integrate on
larger areas at larger radii. A more accurate background level for
these profiles is recomputed on the stacked image between 100 and 150
pixels. The complete covariance matrix of the extracted
background-subtracted SB profile is evaluated using
the jackknife technique.

\subsection{Subtracting the effective PSF}\label{PSFsub}

To isolate the excess light due to intervening absorbers, the
mean point spread function (PSF) must be subtracted.
The PSF depends on several parameters, such as the time of the observation,
the position on a given camera column, and the color of the object. In
order to build an effective PSF which is representative of the
flux-averaged PSF of our QSOs extended to large angular distances, we
stack stars selected to match each individual QSO. We first require
stars to be observed in the same observing run and same
camera column as the QSO. These stars are then ranked to minimize a
combination of the following quantities:
(i) the difference of Gaussian FWHM with respect to the QSO; (ii) the
angular separation from the QSO; (iii) the differences in colors 
with respect to those of the QSO. To map the PSF with reasonable S/N at
large distances, bright unsaturated stars are also preferred ($r<17.0$
mag is always required). For each QSO in each band, the star that best
fulfills all these criteria is chosen.
To obtain the correct flux-weighted average
PSF, the intensity of each star is rescaled to the same intensity of
the corresponding QSO before stacking.

\section{Results}\label{section_results}

\begin{figure}
\plotone{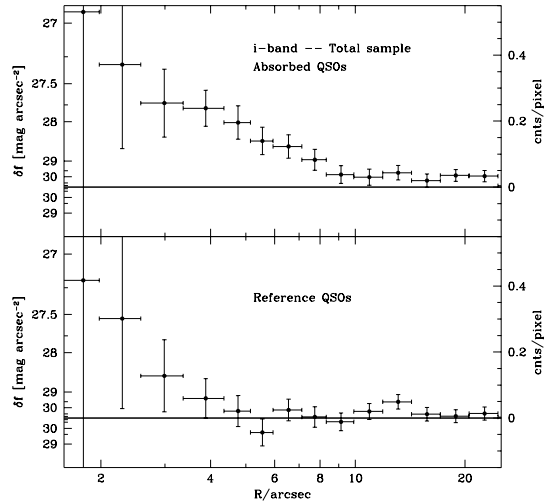}
\caption{$i$-band QSO$-$PSF residuals as a function of 
radius $R$ for the
absorbed QSOs (upper panel) and the \emph{reference} unabsorbed QSOs
(lower panel).  The y-axis scale is linear in flux intensity, as
indicated on the right-side (arbitrary count units per pixel). On the
left the corresponding levels of the surface brightness excess or
deficit are reported in mag arcsec$^{-2}$. The horizontal solid line
indicates the background level.  Error bars are computed with a
jackknife algorithm.\label{plot_res}}
\end{figure}

In Fig. \ref{plot_res} we show the SB residuals measured in the
$i$-band after the PSF subtraction for the sample of 681 QSOs with
absorbers (upper panel) and for the reference QSO sample (lower
panel). The stacked PSFs are normalized with respect to the QSO
profiles within an aperture of 4
pixels (1.6\arcsec). The vertical scale is linear in flux intensity
and the horizontal solid line is the background level.  The left
vertical axis displays the surface brightness in mag arcsec$^{-2}$ (in
excess or deficit).  Error bars are computed from the jackknife
covariance matrix and include the budget from the background
uncertainty. Central normalization uncertainties are found to be
negligible.  
We note that within $\sim 4$\arcsec~both profiles display SB excesses:
these systematic departures from the stellar PSF
is due to reduced efficiency of source de-blending close to 
the QSO and, partly, to the presence of the QSO host galaxy. 
However, we see very clearly that from 3\arcsec~outwards
the SB excess is significantly larger for absorbed QSOs. In particular,
between 4 and 8\arcsec~the absorbed QSOs display positive SB excesses
that are significant at 2 to 4 $\sigma$ level each, while the profile 
of the reference QSOs is fully consistent with the background level.
Similar results are found for the $r$- and $g$-band as well.
Note that galaxies which produce undetected absorbing systems
at different redshifts are expected to be present with the same 
statistically uniform surface density as in the background. Therefore the
SB excess in the absorbed with respect to the reference QSOs
are solely due to galaxies related to the selected absorption systems.
\footnote{We cannot exclude, however, that more than one galaxy per
absorber contributes to the signal. This might be the case if absorbing
galaxies preferentially reside in groups.}

In Table \ref{integrated_mags} we report the fluxes obtained by 
integrating the PSF-subtracted SB within 1.6 and 10\arcsec~($\sim 80$
kpc in the $z_{\mathrm{abs}}$ interval considered)
for the absorbed QSOs in $g$-, $r$-, and $i$-band (columns 1, 3, and 5,
respectively, in mag). The attached 
random errors (1 $\sigma$) are computed by propagating the errors on
the residual SB profiles using the complete jackknife
covariance matrix.
In columns 2, 4, and 6 we report the systematic shifts that must be
applied to the raw magnitudes integrated on the absorbers samples
to correct for the bias in the inner 5\arcsec, as measured from
the reference sample. Corresponding 1 $\sigma$ upper and lower limits
for this correction are also attached.
Measurements are reported for the whole sample (681 QSOs) and for three
subsamples of absorber systems binned in $z_{\mathrm{abs}}$:
$0.37<z_{\mathrm{abs}}\leq0.82$,
$0.82<z_{\mathrm{abs}}\leq1.26$ and $1.26<z_{\mathrm{abs}}\leq2.3$.
Assuming the best estimated value for the systematic corrections,
for the complete sample
we have $\sim 4 \sigma$ detection in $g$- and $r$-band, and $6 \sigma$ 
in $i$. However, even with the most conservative corrections our detections
are significant at 2.9 ($g$ and $r$) and 5.3 $\sigma$ ($i$).
As expected, looking at the three redshift bins, we note that the signal
is clearly dominated by the low-$z_{\mathrm{abs}}$ one. For this subsample
the significance is $2.6~\sigma$ in $g$, 5.6 in $r$, and 9.4 in $i$. 
In the two higher redshift
bins the detections in individual bands are not statistically significant,
however the best estimates we derive are consistent with   
the dimming determined by the larger distance.
\begin{deluxetable*}{lrrrrrr}
\tabletypesize{\scriptsize}
\tablecaption{Integrated magnitudes of QSO absorbers (see Section \ref{section_results}).\label{integrated_mags}}
\tablehead{
\colhead{Sample}&\multicolumn{2}{c}{g}&\multicolumn{2}{c}{r}&\multicolumn{2}{c}{i}\\
&\colhead{m($<10$\arcsec)}&\colhead{$\Delta$m sys.}
&\colhead{m($<10$\arcsec)}&\colhead{$\Delta$m sys.}
&\colhead{m($<10$\arcsec)}&\colhead{$\Delta$m sys.}\\
&\colhead{(1)}&\colhead{(2)}
&\colhead{(3)}&\colhead{(4)}
&\colhead{(5)}&\colhead{(6)}
}
\startdata
All                      & $23.90^{-0.18}_{+0.22}$ & $+0.13^{-0.13}_{+0.24}$ & $22.98^{-0.14}_{+0.16}$ & $+0.39^{-0.12}_{+0.15}$ & $22.31^{-0.11}_{+0.12}$ &  $+0.28^{-0.08}_{+0.08}$\\
$z_{\mathrm{abs}}\leq0.82$        & $23.76^{-0.28}_{+0.38}$ & $+0.00^{-0.00}_{+0.12}$ & $22.25^{-0.13}_{+0.16}$ & $+0.10^{-0.10}_{+0.11}$ & $21.40^{-0.08}_{+0.09}$ &  $+0.15^{-0.06}_{+0.05}$\\
$0.82 < z_{\mathrm{abs}}\leq1.26$ & $24.55^{-0.48}_{+0.90}$ & $+0.17^{-0.17}_{+0.68}$ & $23.51^{-0.39}_{+0.59}$ & $+1.03^{-0.48}_{+0.90}$ & $23.63^{-0.47}_{+0.84}$ &  $+0.74^{-0.53}_{+1.10}$\\
$z_{\mathrm{abs}}>1.26$           & $23.63^{-0.24}_{+0.30}$ & $+0.45^{-0.33}_{+0.78}$ & $23.69^{-0.32}_{+0.47}$ & $+1.21^{-0.66}_{+1.96}$ & $23.09^{-0.36}_{+0.52}$ &  $+0.72^{-0.35}_{+0.51}$
\enddata
\end{deluxetable*}

We now focus on the low-$z_{\mathrm{abs}}$ subsample to derive more
physical information about the absorbing galaxies.
In order to check whether the signal is dominated by a few anomalously bright
objects, we have computed the luminosity function (LF) of
all resolved sources below the masking threshold around the absorbed QSOs
and statistically subtracted the analogous LF in the \emph{reference} QSOs
fields. We find that less than
5\% of the flux is contributed by galaxies that are more
than 2 mag brighter than the average values given in Table \ref{integrated_mags},
while 80\% is contributed by objects brighter than the average. 
The properties we derive are thus well representative of the \emph{luminosity
weighted} average galaxy emission associated to absorbing
systems.\\
From the photometric data
presented above we derive colors within 10\arcsec~aperture:
$g-r=1.41\pm0.35^{+0.16}_{-0.11}$ and $r-i=0.80\pm0.16^{+0.13}_{-0.11}$,
(the first uncertainty is random error on the direct measure, the second 
accounts for errors on the systematic correction).
By neglecting the distribution of absorbers in redshift and assuming
the median redshift of the sample, the comparison
with BC03 models of solar-metallicity dust-free single stellar populations
indicates that the luminosity-weighted average SED of MgII absorbing 
galaxies is dominated by fairly
young stellar populations ($\sim 500$ Myr old), with a rest-frame
$r$-band absolute magnitude of $\sim -21$. The
inferred rest-frame $B-K$ value is about 
2.5 mag, which is compatible with although slightly bluer than the value
found by \citet{Steidel94} using direct imaging and
spectroscopic follow up. A more sophisticated analysis, in which we compare
the observed colors with the $z$-folding of UV-optical galaxy templates 
\citep[from][]{kinney+96} normalized to a fixed $M_i$,
confirms that the ``average'' absorbing galaxy has SED similar to local
intermediate spirals (Sb-c). More details will be presented in
a forthcoming paper.


\section{Conclusions and outlook}
\label{section_conclusions}

In this \emph{Letter} we have used a sample of 681 SDSS-EDR QSOs with
MgII absorption lines ($W_0>0.8$~\AA) to demonstrate the
feasibility of studying the statistical photometric properties of
galaxies associated to MgII absorbing clouds with the SDSS. By
stacking images of quasars and subtracting the PSF, we have obtained
significant detections (up to 9.4 $\sigma$ in $i$-band for the low-redshift
subsample) of surface brightness excesses around QSOs with
strong MgII absorbers in the $g$, $r$ and $i$ bands.
Systematic biases are measured and removed with 10 to $\sim20$\% accuracy
(for the total and low-$z_{\mathrm{abs}}$ samples) by using a control sample
of QSOs without strong absorbers.

As expected, the
measured excess light decreases as a function of absorber redshift.
At $z\sim0.6$, the mean colors ($g-r=1.4$ and
$r-i=0.8$) indicate the presence of relatively young stellar populations and
are consistent with the typical absorbing galaxy resembling a local Sb-c
spiral. The average luminosity per absorber within $\sim 80$ kpc impact
parameter is similar to $L_*$.
Both results are compatible with previous finding by \citet{Steidel94}.
A detailed analysis of impact parameter distributions and constraints
on the corresponding stellar populations will be presented in a
forthcoming paper.

Previously, the largest photometric study of MgII absorbing galaxies
was limited to $\sim60$ low redshift systems \citep{Steidel94} due to
the expensive deep imaging and spectroscopic follow up. Our technique
allows the study of much larger samples: in this \emph{Letter} we have
analyzed a sample more than ten times larger, and it will be soon
applied to the SDSS absorber database (York et al., in preparation),
i.e. another order of magnitude larger. 
Although the statistical nature of our method makes it very difficult to
study the real intrinsic variance of the absorbing galaxy properties,
it can provide robust statistical results and also minimize the problem
of confusion due to chance line-of-sight superposition of galaxies. 
Moreover this technique can be applied to
different absorption line species (FeI, FeII, CIV, DLAs, etc.) 
to measure their impact parameters and investigate their spatial
distribution in the gas around galaxies. Finally, it can also
be used to investigate the properties of intrinsic absorber
systems and quasar host galaxies.
\acknowledgements
We thank the anonymous referee for comments that have greatly improved
the paper, Jim Gunn and Masataka Fukugita for useful
discussions. B.M. acknowledge a support of the Florence Gould
Foundation. Funding for the creation and distribution of the SDSS
Archive has been provided by the Alfred P. Sloan Foundation, the
Participating Institutions, the National Aeronautics and Space
Administration, the National Science Foundation, the U.S. Department
of Energy, the Japanese Monbukagakusho, and the Max Planck Society.

\end{document}